\documentclass[twocolumn]{autart}    

\usepackage{graphicx}          
\usepackage{graphics}
\usepackage{epsfig}

\usepackage{amssymb}

\usepackage{enumerate}
\usepackage{color}
\usepackage{cases}
\newtheorem{assumption}{Assumption}
\newtheorem{lemma}{Lemma}

\newtheorem{theorem}{Theorem}
\newtheorem{remark}{Remark}

\begin{document}

\begin{frontmatter}
\title{Distributed Average Tracking for Lipschitz-Type
Nonlinear Dynamical Systems} 
\thanks[footnotemark]{Corresponding author }
\author[Nwpu] {Yu Zhao \thanksref{footnotemark}}\ead{yuzhao5977@gmail.com},
\author[Nwpu] {Yongfang Liu},
\address[Nwpu]{School of Automation, Northwestern Polytechnical University, Xi'an Shaanxi, 710129, China}



\begin{abstract}                          
In this paper, a distributed average tracking
 problem is studied for Lipschitz-type
nonlinear dynamical systems. The objective is
to design distributed average tracking algorithms for locally interactive agents to track the average of multiple reference signals. Here, in both the agents' and the reference signals' dynamics, there is a nonlinear term satisfying the Lipschitz-type condition. Three types of distributed average tracking algorithms are designed. First, based on state-dependent-gain designing approaches,  a robust distributed average tracking algorithm is developed to solve distributed average tracking problems without requiring the same initial condition. Second, by using a gain adaption
scheme, an adaptive distributed average tracking algorithm is proposed in this paper to remove the requirement that the Lipschitz constant is known for agents. Third, to reduce chattering and make the algorithms easier to implement,
a continuous distributed average tracking algorithm based on a time-varying
boundary layer is further designed as a continuous
approximation of the previous discontinuous distributed average tracking algorithms.
\end{abstract}

\end{frontmatter}
 
\section{Introduction}
{In the past two decades, there {have} been lots of interests in the distributed cooperative control \cite{OlfatiSaber1}, \cite{Ren:07}, \cite{Hong:08}, \cite{Cao:12}, \cite{Litac}, \cite{Tuna:2008}, \cite{Zhang:11}, \cite{Liu:15}, \cite{Liu:16}, \cite{Zhao:16}, \cite{Zhao:15}, \cite{Zhao:15scl}, and \cite{Ji}, for multi-agent systems due to its {potential} applications in formation flying, path planning and so forth.
Distributed average tracking, as a generalization of consensus and cooperative tracking {problems}, has received increasing attentions and been applied in many different perspectives, such as distributed sensor networks \cite{Spanos2}, \cite{Bai2011} and distributed coordination \cite{Yang}, \cite{Sun}. For practical applications, distributed average tracking should be investigated for signals modeled by more and more complex dynamical systems.}

The objective of distributed average tracking problems is to design a distributed algorithm for multi-agent systems to track the average of multiple reference signals.
The motivation of this problem comes from the coordinated tracking for multiple camera systems. Spurred by the pioneering works in \cite{Spanos1}, and \cite{Freeman} on the distributed average tracking via linear algorithms, { real applications of related results can be found} in distributed sensor fusion \cite{Spanos2}, \cite{Bai2011}, and formation control \cite{Yang}. In \cite{Bai}, distributed average tracking problems were investigated by considering the robustness to initial errors in algorithms. The above-mentioned results are important for scientific researchers to build up a general framework to investigate this topic. {However, a common assumption in the above works is that the multiple reference signals are constants \cite{Freeman} or achieving to values  \cite{Spanos1}.} In practical applications, reference signals may be produced by more general dynamics. {For this reason,} a class of nonlinear algorithms were designed in \cite{Nosrati:12} to track multiple reference signals with bounded deviations. Then, based on non-smooth control {approaches}, a couple of distributed algorithms were proposed in \cite{Chengfei:12} and \cite{Chengfei:13} for agents to track arbitrary time-varying reference signals with bounded {deviations} and {bounded second deviations}, respectively. Using discontinuous algorithms, further, \cite{Zhaoyuicca} studied the distributed average tracking problems for multiple signals generated by linear dynamics.

Motivated by the above mentioned observations, this paper is devoted to solving the distributed average tracking problem for Lipschitz-type
nonlinear dynamical systems. Three DAT algorithms are proposed in this paper. First of all, based on relative states of neighboring agents, a class of distributed discontinuous DAT algorithms are proposed with robustness to initial conditions.  {Then, a novel class of distributed algorithms with adaptive coupling strengths are designed by utilizing an adaptive control technique.}
{Different from \cite{Chengfei:12}, \cite{Chengfei:13} and \cite{Zhaoyuicca}}, the proposed algorithms are based on node adaptive lows. Further, a class of continuous algorithms are given to reduce  chattering.
{Compared with the above existing results, the contributions of this paper are three-fold. First, main results of this paper extend the dynamics of the reference signals and agents from linear systems \cite{Chengfei:12} and \cite{Chengfei:13} to nonlinear systems, which can describe more complex dynamics.   Second, by using adaptive control approaches, the requirements of all  global information  {are} removed, which greatly reduce the computational complexity for large-scale networks.
Third, compared with existing results in \cite{Zhaoyuicca}, new continuous algorithms are redesigned via the boundary layer concept to reduce { the } chattering phenomenon. Continuous algorithms in this paper is more appropriate for real engineering applications.}

\emph{Notations}: Let $R^n$ and $R^{n\times n}$ be sets of real numbers and real matrices, respectively. $I_n$ represents the identity matrix of dimension $n$. Denote by $\mathbf{1}$ a column vector with
all entries equal to one. The matrix inequality $A> (\geq) B$ means that
$A-B$ is positive (semi-) definite. Denote by $A\otimes B$ the Kronecker product
of matrices $A$ and $B$. For a vector $x=(x_1,x_2,\cdots,x_n)^T\in R^n$, let $\|x\|$
denote the 2-norm of $x$, $h(x)=\frac{x}{\|x\|}$, $h_\varepsilon(x)=\frac{x}{\|x\|+\varepsilon e^{-ct}}$. For a set $V$, $|V|$ represents the number of elements in $V$.

\section{Preliminaries}

\subsection{Graph Theory}
An undirected (simple) graph $\mathcal{G}$ is specified by a vertex set $\mathcal{V}$ and an edge set $\mathcal{E}$ whose elements characterize the incidence relations between distinct pairs of $\mathcal{V}$. The notation $i\sim j$ is used to denote that node $i$ is connected to node $j$, or equivalently, $(i, j)\in \mathcal{E}$. We make use of the $|\mathcal{V}|\times|\mathcal{E}|$ incidence matrix, $D(\mathcal{G})$, for a graph with an arbitrary orientation, i.e., a graph whose edges have a head (a terminal node) and a tail (an initial node). The columns of $D(\mathcal{G})$ are then indexed by the edge set, and the $i$th row entry takes the value $1$ if it is the initial node of the corresponding edge, $-1$ if it is the terminal node, and zero otherwise. The diagonal matrix $\Delta(\mathcal{G})$ of the graph contains the degree of each vertex on its diagonal. The adjacency matrix, $A(\mathcal{G})$, is the $|\mathcal{V}|\times|\mathcal{V}|$ symmetric matrix with zero in the diagonal and one in the $(i,j)$th position if node $i$ is adjacent to node $j$. The graph Laplacian  \cite{GraphTheory} of $\mathcal{G}$, $L:= {\frac{1}{2}}D(\mathcal{G})D(\mathcal{G})^T=\Delta(\mathcal{G})-A(\mathcal{G})$,
is a rank deficient positive semi-definite matrix.

An undirected path between node $i_1$ and node $i_s$ on undirected graph means a sequence of ordered undirected edges with the form $(i_k; i_{k+1}), k = 1, \cdots, s-1$.
A graph $\mathcal{G}$ is said to be connected if there exists a path between each pair of distinct nodes.

\begin{assumption}\label{ass}
Graph $\mathcal{G}$ is undirected and connected.
\end{assumption}

\begin{lemma} \label{lemma1}\cite{GraphTheory}
Under Assumption \ref{ass}, zero is a simple eigenvalue of $L$ with
$\mathbf{1}$ as an eigenvector and all the other eigenvalues are positive. Moreover, the
smallest nonzero eigenvalue $\lambda_2$ of L satisfies $\lambda_2=\min\limits_{ x\neq 0, \mathbf{1}^Tx=0 } \frac{x^TLx}{x^Tx}$.
\end{lemma}

\section{Main results}
\subsection{Robust distributed average tracking algorithms design}
Consider a multi-agent system consisting of $N$ physical agents described by the following nonlinear
dynamics
\begin{eqnarray}\label{agent}
\dot{x}_i(t)=Ax_i(t)+Bf(x_i,t)+Bu_i,
\end{eqnarray}
where $A\in R^{n\times n}$ and $B\in R^{n\times p}$ both are constant matrices with compatible dimensions, $x_i(t)\in R^{n}$ and $u_i(t)\in R^{p}$ is the state and control input
of the $i$th agent, respectively, and $f: R^{n}\times R^+\to R^{p}$ is a nonlinear function.
Suppose that there is  a time-varying reference signal, $r_i(t)\in R^n, i=1,2,\cdots, N$, which generated by  the following Lipschitz-type nonlinear dynamical systems:
\begin{eqnarray}\label{L referencesignals}
\dot{r}_i(t)=Ar_i(t)+Bf(r_i,t),
\end{eqnarray}
where {$r_i(t)\in R^{n}$} is the state of the $i$th reference signal.

It is assumed that agent $i$ has access to $r_i(t)$, and agent $i$ can obtain the relative information from its neighbors denoted by $\mathcal{N}_i$.
 {\begin{assumption}\label{ass2}
$(A,B)$ is stabilizable.
\end{assumption}}
The main objective of this paper is to design a class of distributed controller $u_i(t)$ for physical agent $i$ in (\ref{agent}) to track the average of multiple reference signals $r_i(t)$ generated by the general nonlinear dynamics  (\ref{L referencesignals}), i.e., $$\lim_{t\rightarrow \infty}\bigg(x_i(t)-\frac{1}{N}\sum_{i=1}^Nr_i(t)\bigg)=0,$$
where each agent has only local interaction with its neighbors.
{\assumption{\label{assumpf} For  $\forall \theta_i(t) \in R^{n}$, $i=1,2$ and $\forall t>0$, the nonlinear function $f: R^{n}\times R^+\to R^{p}$ satisfies a Lipschitz-type condition:  $\|f(\theta_1,t)-f(\theta_2,t)\|\leq \gamma \|\theta_1-\theta_2\|$, where $\gamma\in R^+$ and  $f(0,t)=0$.}}

As it was mentioned, there are many applications that the
physical agents should track a time varying trajectory, where
each agent has an incomplete copy of this trajectory. While, the
physical agents and reference trajectory might be described by
more complicated dynamics rather than the linear
dynamics in real applications. Therefore, we consider a more
general group of physical agents, where the nonlinear function
$f(\cdot, t)$ in their dynamics satisfies the Lipschitz-type condition.

Therefore, a distributed average tracking controller algorithm is designed as
\begin{eqnarray}\label{control input}
u_i(t) &=&K_1(p_i(t)-r_i(t))+K_2\tilde{x}_i(t)
+\mu\phi_ih[K_2\tilde{x}_i(t)]\nonumber\\
&&+\alpha\vartheta_iBh\bigg(\sum_{j\in \mathcal{N}_i} K_1(p_i(t)-p_j(t))\bigg),
\end{eqnarray}
with a distributed average tracking filter algorithm is proposed as follows:
\begin{eqnarray}\label{L distributed control algorithm}
{p}_i(t) &=& s_i(t)+r_i(t),\nonumber \\
\dot{s}_i(t) &=&As_i(t)+BK_1(p_i(t)-r_i(t))\nonumber\\
&&+\alpha\vartheta_iBh\bigg(\sum_{j\in \mathcal{N}_i} K_1(p_i(t)-p_j(t))\bigg),
\end{eqnarray}
where $\tilde{x}_i(t)=x_i(t)-p_i(t)$, $s_i(t), \;i=1,2,\cdots,N$, are the states of the DAT algorithm, $\phi_i=\|x_i(t)-r_i(t)\|+\nu$, and $\vartheta_i=\|r_i(t)\|+\beta$ state-dependent time-varying parameters,  $\mu$, $\nu$, $\alpha$ and $\beta$ constant parameters, $K_1$ and $K_2$ control
gain matrices, respectively, to be determined.

Then, using the controller (\ref{control input}) for
(\ref{agent}), one gets the tracking error system
\begin{eqnarray}\label{eagents_u}
\dot{\tilde{x}}_i(t)&=&(A+BK_2)\tilde{x}_i(t)+B(f(x_i,t)-f(r_i,t))\nonumber\\
&&+\mu\phi_iBh[K_2\tilde{x}_i(t)].
\end{eqnarray}
Following from (\ref{L referencesignals}) and (\ref{L distributed control algorithm}), one gets
\begin{eqnarray}\label{L closedloop}
\dot{p}_i(t)&=&(A+BK_1)p_i(t)-BK_1r_i(t)+Bf(r_i,t)\nonumber\\
&+&\alpha\vartheta_iBh\bigg(\sum_{j\in \mathcal{N}_i} K_1(p_i(t)-p_j(t))\bigg).
\end{eqnarray}
Let $\tilde{x}(t)=(\tilde{x}_1^T(t),\tilde{x}_2^T(t),\cdots,\tilde{x}_N^T(t))^T$, $p(t)=(p_1^T(t),\\
p_2^T(t),\cdots,p_N^T(t))^T$, $r(t)=(r_1^T(t),
r_2^T(t),\cdots,r_N^T(t))^T$, $\Phi=\mathrm{diag}(\phi_1,\phi_2,
\cdots,\phi_N)$, $\Theta=\mathrm{diag}(\vartheta_1,\vartheta_2,
\cdots,\vartheta_N)$, $F(r,t)=(f^T(r_1,t),f^T(r_2,t),
\cdots,f^T(r_N,t))^T$, and $F(x,t)=(f^T(x_1,t),f^T(x_2,t),\cdots,f^T(x_N,t))^T$. In matrix form, one obtains the closed-loop system as follows:
\begin{eqnarray}\label{Meagents_u}
\dot{\tilde{x}}(t)&=&(I\otimes(A+BK_2))\tilde{x}(t)
+(I\otimes B)(F(x,t)-F(r,t))\nonumber\\
&&+\mu (\Phi\otimes B) H[(I\otimes K_2)\tilde{x}(t)],
\end{eqnarray}
with
\begin{eqnarray}\label{LM closedloop}
\dot{p}(t)&=&(I\otimes (A+BK_1))p(t)-(I\otimes BK_1)r(t)\nonumber\\
&+&(I\otimes B)F(r,t)+\alpha(\Theta\otimes B) H((L\otimes K_1)p(t)),
\end{eqnarray}
where $$H((I\otimes K_2)\tilde{x}(t))=\left(
                               \begin{array}{c}
                                 h(K_2\tilde{x}_1(t)) \\
                                 \vdots \\
                                 h(K_2\tilde{x}_N(t)) \\
                               \end{array}
                             \right),
$$
and
$$H((L\otimes K_1)p(t))=\left(
                               \begin{array}{c}
                                 h\bigg(\sum_{j\in \mathcal{N}_1} K_2(p_1(t)-p_j(t))\bigg) \\
                                 \vdots \\
                                 h\bigg(\sum_{j\in \mathcal{N}_N} K_2(p_N(t)-p_j(t))\bigg) \\
                               \end{array}
                             \right).
$$

{Define $M=I_N-\frac{1}{N}\mathbf{1}\mathbf{1}^T$. Then $M$ satisfies following properties: Firstly, it is easy to see that $0$ is a simple eigenvalue
of $M$ with $\mathbf{1}$ as {the} corresponding right eigenvector and $1$ is the other eigenvalue with multiplicity $N-1$, i.e., $M\mathbf{1}=\mathbf{1}^TM=0$. Secondly, since $L^T=L$, one has $LM=L(I_N-\frac{1}{N}\mathbf{1}\mathbf{1}^T)=L-\frac{1}{N}L\mathbf{1}\mathbf{1}^T=L=L-\frac{1}{N}\mathbf{1}\mathbf{1}^TL=(I_N-\frac{1}{N}\mathbf{1}\mathbf{1}^T)L=
ML$. Finally, $M^2=M(I_N-\frac{1}{N}\mathbf{1}\mathbf{1}^T)=M-\frac{1}{N}M\mathbf{1}\mathbf{1}^T=M$.
}

Define $\xi(t)=(M\otimes I)p(t)$, where $\xi(t)=(\xi_1^T(t),\xi_2^T(t),\cdots,\\
\xi_N^T(t))^T$.
Then, it follows that $\xi(t)= 0$ if and only if $p_1(t)=p_2(t)=\cdots=p_N(t)$. Therefore, the consensus problem of (\ref{L closedloop}) is solved if and only if $\xi(t)$ asymptotically converges to zero. Hereafter, we refer to $\xi(t)$ as the consensus error. By
noting that $LM = L=ML$, it is not difficult to obtain from (\ref{LM closedloop}) that the consensus error $\xi(t)$ satisfies
\begin{eqnarray}\label{LME closedloop}
\dot{\xi}(t)&=&(M\otimes (A+BK_1))\xi(t)-(M\otimes BK_1)r(t)\nonumber\\
\!\!&+&\alpha(M\Theta\otimes B) H(L\otimes K_1)\xi(t)+(M\otimes B)F(r,t).
\end{eqnarray}

\textbf{Algorithm 1:} {Under Assumptions 1 and 2,} for multiple reference signals in (\ref{L referencesignals}), the distributed average tracking algorithms (\ref{L distributed control algorithm}) and (\ref{control input}) can be constructed as follows
\begin{enumerate}
  \item Solve the following algebraic Ricatti equations (AREs):
\begin{eqnarray}\label{LMI}
P_iA+A^TP_i-P_iBB^TP_i+Q_i=0,
\end{eqnarray}
with $Q_i>0$ to obtain matrices  $P_i>0$, where $i=1,2$. Then, choose $K_i=-B^TP_i,\; i=1,2$.
  \item Choose the parameters $\alpha\geq \gamma+\|B^TP_1\|$, $\beta>0$ $\mu\geq \gamma$ and $\nu>0$.
\end{enumerate}


\begin{theorem}\label{L theorem1f}
Under Assumptions 1-3, by using the distributed average tracking controller algorithm (\ref{control input}) with the distributed average tracking filter algorithm (\ref{L distributed control algorithm}), the state $x_i(t)$ in (\ref{agent}) will track the average of multiple reference signals $r_i(t),\;i=1,2,\cdots, N$, generated by the Lipschitz-type nonlinear dynamical systems (\ref{L referencesignals}) if the parameters $\alpha$, $\beta$, $\mu$, $\nu$ and the feedback gains $K_i, i=1,2,$ are designed by Algorithm 1.
\end{theorem}
\textbf{Proof}:
The proof contains three steps. First, it is proved that for the $i$th agent, $\lim_{t\rightarrow \infty}\bigg(p_i(t)-\frac{1}{N}\sum_{k=1}^{N}p_k(t)\bigg)
=0
$.
Consider the Lyapunov function candidate
\begin{eqnarray}\label{LV 1}
V_1(t)= \xi^T(L\otimes P_1)\xi.
\end{eqnarray}
By the definition of $\xi(t)$, it is easy to see that $(\mathbf{1}^T\otimes I)\xi = 0$.
For the connected graph $\mathcal{G}$, it then follows from Lemma \ref{lemma1} that
\begin{eqnarray}\label{gLV 1}
V_1(t)\geq \lambda_2\lambda_{\min}(P_1)\|\xi\|^2,
\end{eqnarray}
where $\lambda_{\min}(P_1)$ is the smallest eigenvalue of the positive matrix $P_1$.
The time derivative of $V_1$ along (\ref{LME closedloop}) can be obtained as follows
\begin{eqnarray}\label{dLV1}
\dot{ {V}}_1&=& \dot{\xi}^T(L\otimes P_1)\xi+ \xi^T(L\otimes P_1)\dot{\xi} \nonumber\\
&=& \xi^T(M\otimes (A+BK_1)^T)(L\otimes P_1)\xi\nonumber\\
&&+\xi^T(L\otimes P_1)(M\otimes (A+BK_1))\xi\nonumber\\
&&-2\xi^T(L\otimes P_1)(M\otimes BK_1)r(t)\nonumber\\
&&+2 \alpha\xi^T(L\otimes P_1)(M\Theta\otimes B) H(L\otimes K_1)\xi(t) \nonumber\\
&&+2\xi^T(L\otimes P_1)(M\otimes B)F(r,t).
\end{eqnarray}
Substituting $K_1=-B^TP_1$ into (\ref{dLV1}), it follows from the fact $LM=ML=L$ and Assumption \ref{assumpf} that
\begin{eqnarray}\label{ddLV1}
\dot{ {V}}_1
&=& \xi^T[L\otimes (A^TP_1+P_1A)-2(L\otimes P_1BB^TP_1)]\xi\nonumber\\
&&+2\xi^T(L\otimes P_1BB^TP_1)r(t)\nonumber\\
&&-2 \alpha\xi^T(L\Theta\otimes PB) H[(L\otimes B^TP_1)\xi] \nonumber\\
&&+2\xi^T(L\otimes P_1B)F(r,t)\nonumber\\
&=& \xi^T[L\otimes (A^TP_1+P_1A)-2(L\otimes P_1BB^TP_1)]\xi\nonumber\\
&&+2\sum_{i=1}^N \bigg(\sum_{j\in \mathcal{N}_i}[B^TP_1(\xi_i(t)-\xi_j(t))]\bigg)^TB^TP_1r_i\nonumber\\
&&-2 \alpha\sum_{i=1}^N\vartheta_i\bigg(\sum_{j\in \mathcal{N}_i}[B^TP_1(\xi_i(t)-\xi_j(t))]\bigg)^T\nonumber\\
&&h\bigg(\sum_{j\in \mathcal{N}_i}[B^TP_1(\xi_i(t)-\xi_j(t))]\bigg)
\nonumber\\
&&+2\sum_{i=1}^N \bigg(\sum_{j\in \mathcal{N}_i}[B^TP_1(\xi_i(t)-\xi_j(t))]\bigg)^T[f(r_i,t)-f(0,t)]\nonumber\\
&\leq& \xi^T[L\otimes (A^TP_1+P_1A)-2(L\otimes P_1BB^TP_1)]\xi\nonumber\\
&&+2\bigg\|\sum_{i=1}^N \bigg(\sum_{j\in \mathcal{N}_i}[B^TP_1(\xi_i(t)-\xi_j(t))]\bigg)^T\bigg\| \|B^TP_1r_i\| \nonumber\\
&&-2 \alpha\sum_{i=1}^N\vartheta_i\bigg\|\sum_{j\in \mathcal{N}_i}[B^TP_1(\xi_i(t)-\xi_j(t))]\bigg\| \nonumber\\
&&+2\sum_{i=1}^N \bigg\|\sum_{j\in \mathcal{N}_i}[B^TP_1(\xi_i(t)-\xi_j(t))]\bigg\|\|f(r_i,t)-f(0,t)\|\nonumber\\
&\leq& \xi^T[L\otimes (A^TP_1+P_1A)-2(L\otimes P_1BB^TP_1)]\xi\nonumber\\
&&-2 \alpha\sum_{i=1}^N\vartheta_i\bigg\|\sum_{j\in \mathcal{N}_i}[B^TP_1(\xi_i(t)-\xi_j(t))]\bigg\|
\nonumber\\
&&+2\sum_{i=1}^N \bigg\|\sum_{j\in \mathcal{N}_i}[B^TP_1(\xi_i(t)-\xi_j(t))]\bigg\|(\gamma+\|B^TP\|_1)\|r_i\|\nonumber\\
&=& \xi^T[L\otimes (A^TP_1+P_1A)-2(L\otimes P_1BB^TP_1)]\xi\nonumber\\
&&-2 \sum_{i=1}^N[(\alpha-\gamma-\|B^TP_1\| )\|r_i\| +\alpha\beta]\nonumber\\
&&\bigg\|\sum_{j\in \mathcal{N}_i}[B^TP_1(\xi_i(t)-\xi_j(t))]\bigg\|
.
\end{eqnarray}
Since $\alpha>\gamma+\|B^TP_1\|, \beta>0$, one has
\begin{eqnarray}\label{ddddLV1}
\dot{ {V}}_1
&\leq&\xi^T(L {\otimes} (P_1A{+}A^TP_1-2P_1BB^TP_1))\xi
\nonumber\\
&\leq&\lambda_2\xi^T(I {\otimes} (P_1A{+}A^TP_1-2P_1BB^TP_1))\xi
.
\end{eqnarray}
It follows from (\ref{LMI}) that $P_1A+A^TP_1-P_1BB^TP_1\leq-Q_1$. Therefore, we have
\begin{eqnarray}\label{ddddddLV1}
\dot{ {V}}_1&<&- \eta_1 V_1,
\end{eqnarray}
where $\eta_1=\frac{\lambda_{\min}(Q_1)}{\lambda_{\max}(P_1)}$.
Thus, one has
$$
\lim_{t\rightarrow \infty}\xi_i(t)=\lim_{t\rightarrow \infty}\bigg(p_i(t)-\frac{1}{N}\sum_{k=1}^{N}p_k(t)\bigg)
=0.
$$
Second, it is proved that
$\lim_{t\rightarrow \infty}\bigg(p_i(t)-\frac{1}{N}\sum_{k=1}^{N}r_k(t)\bigg)=0$.
Let $r^*(t)=\frac{1}{N}\sum_{i=1}^Nr_i(t)$. It follows from (\ref{L referencesignals}) that
\begin{eqnarray}\label{AL referencesignals}
\dot{r}^*(t)=Ar^*(t)+\frac{1}{N}B\sum_{i=1}^Nf(r_i(t),t).
\end{eqnarray}
Let $p^*(t)=\frac{1}{N}\sum_{i=1}^Np_i(t)$. It follows from (\ref{L referencesignals}) that
\begin{eqnarray}\label{AX referencesignals}
\dot{p}^*(t)&=&(A+BK_1)p^*(t)-BK_1r^*(t)+\frac{1}{N}B\sum_{i=1}^Nf(r_i(t),t)\nonumber\\
&+&\alpha\sum_{i=1}^N\vartheta_ih\bigg(\sum_{j\in \mathcal{N}_i} K_1(p_i(t)-p_j(t))\bigg).
\end{eqnarray}
Denote $\zeta(t)=p^*(t)-r^*(t)$, one has
\begin{eqnarray}\label{EAL referencesignals}
\dot{\zeta}(t)
&=&\dot{p}^*(t)-\dot{r}^*(t)\nonumber\\
&=&(A+BK_1)p^*(t)-BK_1r^*(t)-Ar^*(t)\nonumber\\
&+&\alpha\sum_{i=1}^N\vartheta_ih\bigg(\sum_{j\in \mathcal{N}_i} K_1(p_i(t)-p_j(t))\bigg)\nonumber\\
&=&(A+BK_1){\zeta}(t)+\omega(t),
\end{eqnarray}
where $\omega(t)=\alpha\sum_{i=1}^N\vartheta_ih\bigg(\sum_{j\in \mathcal{N}_i} K_1(p_i(t)-p_j(t))\bigg)$.
We then use input-to-state stability to analyze the system (\ref{EAL referencesignals}) by treating the term $\omega(t)$ as the input and $\zeta(t)$ as the states. Since (\ref{LMI}) with $K_1=-B^TP_1$, one has $A+BK_1$ is Hurwitz. Thus, the system (\ref{EAL referencesignals})
with zero input is exponentially stable and hence input-to-state stable. Since $\lim_{t\rightarrow \infty}\bigg(p_i(t)-\frac{1}{N}\sum_{k=1}^{N}p_k(t)\bigg)
=0$. One has $\lim_{t\rightarrow \infty}\omega(t)=0$. Thus, it follows that $\lim_{t\rightarrow \infty}\zeta(t)=0$, which implies that
$\lim_{t\rightarrow \infty}\bigg(\frac{1}{N}\sum_{i=1}^Np_i(t)-\frac{1}{N}\sum_{i=1}^Nr_i(t)\bigg)=0$.
Therefore, one obtains
$\lim_{t\rightarrow \infty}\bigg(p_i(t)-\frac{1}{N}\sum_{k=1}^{N}r_k(t)\bigg)=\lim_{t\rightarrow \infty}\bigg(p_i(t)-\frac{1}{N}\sum_{i=1}^Np_i(t)\bigg)+\lim_{t\rightarrow \infty}\bigg(\frac{1}{N}\sum_{i=1}^Np_i(t)-\frac{1}{N}\sum_{i=1}^Nr_i(t)\bigg)=0$.
\\
Third, it is proofed that $\lim_{t\rightarrow \infty}\bigg(x_i(t)-\frac{1}{N}\sum_{i=1}^Nr_i(t)\bigg)=0$.
Consider the candidate Lyapunov function
\begin{eqnarray}\label{V2}
V_2=\tilde{x}^T(I\otimes P_2)\tilde{x},
\end{eqnarray}
with $P_2>0$. By taking the derivative of $V_2$ along (\ref{Meagents_u}), one gets
\begin{eqnarray}\label{dV2}
\dot{V}_2&=&\tilde{x}^T(I\otimes ((A+BK_2)^TP_2+P_2(A+BK_2)))\tilde{x}\nonumber\\
&&+2\tilde{x}^T(I\otimes P_2B)(F(x,t)-F(r,t))\nonumber\\
&&+2\mu (\Phi\otimes P_2B) H[(I\otimes K_2)\tilde{x}(t)].
\end{eqnarray}
Using $K_2=-B^TP_2$, one has
\begin{eqnarray}\label{ddV2}
\dot{V}_2&=&\tilde{x}^T(I\otimes (A^TP_2+P_2A-2P_2BB^TP_2))\tilde{x}
\nonumber\\
&&+2\tilde{x}^T(I\otimes P_2B)(F(x,t)-F(r,t))\nonumber\\
&&-2\mu \tilde{x}^T(\Phi\otimes P_2B)  H[(I\otimes B^TP_2)\tilde{x}(t)]\nonumber\\
&=&\tilde{x}^T(I\otimes (A^TP_2+P_2A-2P_2BB^TP_2))\tilde{x}\nonumber\\
&&
+2\sum_{i=1}^N(B^TP_2\tilde{x}_i(t))^T(f(x_i,t)-f(r_i,t))\nonumber\\
&&-2\mu\sum_{i=1}^N\phi_i (B^TP_2\tilde{x}_i(t))^Th(B^TP_2\tilde{x}_i)\nonumber\\
&\leq&\tilde{x}^T(I\otimes (A^TP_2+P_2A-2P_2BB^TP_2))\tilde{x}\nonumber\\
&&
+2\sum_{i=1}^N\|B^TP_2\tilde{x}_i(t)\|\|(f(x_i,t)-f(r_i,t))\|\nonumber\\
&&-2\mu\sum_{i=1}^N\phi_i \|B^TP_2\tilde{x}_i(t)\|
\nonumber\\
&\leq&\tilde{x}^T(I\otimes (A^TP_2+P_2A-2P_2BB^TP_2))\tilde{x}\nonumber\\
&&
+2\sum_{i=1}^N\|B^TP_2\tilde{x}_i(t)\|\gamma\|x_i-r_i\|\nonumber\\
&&-2\mu\sum_{i=1}^N(\|x_i-r_i\|+\nu) \|B^TP_2\tilde{x}_i(t)\|\nonumber\\
&\leq&\tilde{x}^T(I\otimes (A^TP_2+P_2A-2P_2BB^TP_2))\tilde{x}\nonumber\\
&&-2\sum_{i=1}^N((\mu-\gamma)\|x_i-r_i\|+\mu\nu) \|B^TP_2\tilde{x}_i(t)\|.
\end{eqnarray}
Since $\mu\geq \gamma$ and $\nu>0$, one has
\begin{eqnarray}\label{dddV2}
\dot{V}_2
&\leq&\tilde{x}^T(I\otimes (A^TP_2+P_2A-2P_2BB^TP_2))\tilde{x}.
\end{eqnarray}
Using $A^TP_2+P_2A-2P_2BB^TP_2\leq -Q_2$, one has
\begin{eqnarray}\label{dddV2}
\dot{V}_2
&\leq&-\eta_2{V}_2.
\end{eqnarray}
where $\eta_2=\frac{\lambda_{\min}(Q_2)}{\lambda_{\max}(P_2)}$.
Thus, one has $\lim_{t\rightarrow \infty}\bigg(x_i(t)-\frac{1}{N}\sum_{i=1}^Nr_i(t)\bigg)=\lim_{t\rightarrow \infty}(x_i(t)-p_i(t))+\bigg(p_i(t)-\frac{1}{N}\sum_{i=1}^Nr_i(t)\bigg)=0$.
Therefore, the distributed average tracking problem is solved. This completes the proof.

\subsection{Adaptive distributed average tracking algorithms design}
Note that, in above subsection, the proposed distributed average tracking algorithms (\ref{control input}) and (\ref{L distributed control algorithm}) require that the parameters $\alpha$ and $\mu$ satisfy the conditions $\alpha\geq \gamma+\|B^TP_1\|$ and  $\mu\geq \gamma$, which depend the Lipschitz constant $\gamma$. Since  the $\gamma$ is a global information, for a local agent, it becomes difficult to obtain $\gamma$. Therefore, to overcome the global information restriction, we design an adaptive distributed average tracking controller algorithm
\begin{eqnarray}\label{Adaptive control input}
u_i(t) &=&K_1(p_i(t)-r_i(t))+K_2\tilde{x}_i(t)
+\mu_i(t)\phi_ih[K_2\tilde{x}_i(t)]\nonumber\\
&&+\alpha_i(t)\vartheta_iBh\bigg(\sum_{j\in \mathcal{N}_i} K_1(p_i(t)-p_j(t))\bigg),
\end{eqnarray}
and an adaptive distributed average tracking filter algorithm
\begin{eqnarray}\label{Adaptive DAT algorithm}
{p}_i(t) &=& s_i(t)+r_i(t),\nonumber \\
\dot{s}_i(t) &=&As_i(t)+BK_1(p_i(t)-r_i(t))\nonumber\\
&&+\alpha_i(t)\vartheta_iBh\bigg(\sum_{j\in \mathcal{N}_i} K_1(p_i(t)-p_j(t))\bigg),
\end{eqnarray}
with two time-varying parameters $\mu_i(t)$ and $\alpha_i(t)$ satisfying the following adaptive update strategies:
\begin{eqnarray}\label{Adaptive mu}
\dot{\mu}_i(t)=\kappa_i \phi_i\|K_2\widetilde{x}_i(t)\|,
\end{eqnarray}
and \begin{eqnarray}\label{Adaptive alpha}
\dot{\alpha}_i(t)\!\!=\!\!\chi_i \vartheta_i\bigg\|
\sum_{j\in \mathcal{N}_i}K_1(\xi_i(t){-}\xi_j(t))\bigg\|,
\end{eqnarray}
respectively, where $\kappa_i, \chi_i$ are adaptive parameters to be determined.

By substituting adaptive controller (\ref{Adaptive control input}) into
(\ref{agent}), one obtains
\begin{eqnarray}\label{Aeagents_u}
\dot{\tilde{x}}_i(t)&=&(A+BK_2)\tilde{x}_i(t)+B(f(x_i,t)-f(r_i,t))\nonumber\\
&&+\mu_i(t)\phi_iBh[K_2\tilde{x}_i(t)],
\end{eqnarray}
where $\mu_i(t)$ is given by (\ref{Adaptive mu}).
According to (\ref{L referencesignals}) and (\ref{Adaptive DAT algorithm}), one has
\begin{eqnarray}\label{A closedloop}
\dot{p}_i(t)&=&(A+BK_1)p_i(t)-BK_1r_i(t)+Bf(r_i,t)\nonumber\\
&+&\alpha_i(t)\vartheta_iBh\bigg(\sum_{j\in \mathcal{N}_i} K_1(p_i(t)-p_j(t))\bigg),
\end{eqnarray}
where $\alpha_i(t)$ is given by (\ref{Adaptive alpha}).
 
Then, the closed-loop systems in matrix form are obtained,
\begin{eqnarray}\label{AMeagents_u}
\dot{\tilde{x}}(t)&=&(I\otimes(A+BK_2))\tilde{x}(t)
+(I\otimes B)(F(x,t)-F(r,t))\nonumber\\
&&+ (\mu(t)\Phi\otimes B) H[(I\otimes K_2)\tilde{x}(t)],
\end{eqnarray}
with
\begin{eqnarray}\label{AM closedloop}
\dot{\xi}(t)&=&(I\otimes (A+BK_1))\xi(t)-(M\otimes BK_1)r(t)\nonumber\\
&+&(M\otimes B)F(r,t)+(M\alpha(t)\Theta\otimes B) H((L\otimes K_1)\xi(t)),
\end{eqnarray}
where $\mu(t)=\mathrm{diag}(\mu_1(t),\mu_2(t),\cdots, \mu_N(t))$, and $\alpha(t)=\mathrm{diag}(\alpha_1(t),\alpha_2(t),\cdots, \alpha_N(t))$, respectively.

{\assumption{It is assumed that $r_i$ is bounded.}}

\textbf{Algorithm 2:} Under Assumptions 1-4, for multiple reference signals in (\ref{L referencesignals}), the adaptive distributed average tracking algorithms (\ref{Adaptive control input})-(\ref{Adaptive alpha}) is designed by the following two steps:
\begin{enumerate}
  \item Solve the AREs (\ref{LMI}) to obtain $K_i, i=1,2$.
  \item Choose the parameters $\kappa>0, \chi>0, \beta>0,$ and $\nu>0$.
\end{enumerate}

\begin{theorem}\label{A theorem1f}
Under Assumptions 1-4, the adaptive distributed average tracking algorithms (\ref{Adaptive control input})-(\ref{Adaptive alpha}) solve the distributed average tracking problem of the multi-agent system (\ref{agent}) with the reference dynamical system (\ref{L referencesignals}) if the parameters are given by Algorithm 2.
\end{theorem}
\textbf{Proof}: First, consider the following Lyapunov candidate,
\begin{eqnarray}\label{AV3}
V_3 &=&\xi^T(L\otimes P_1)\xi+\sum_{i=1}^N \frac{\widetilde{\alpha}_{i}(t)^2}
{\chi_i},
\end{eqnarray}
where $\widetilde{\alpha}_{i}(t)= \alpha_{i}(t){-} {\alpha}$.
As proved in Theorem 1, the derivation of (\ref{AV3}) along (\ref{AM closedloop}) and (\ref{Adaptive alpha}) is given by
\begin{eqnarray}\label{AddLV1}
\dot{ {V}}_3
&\leq& \xi^T[L\otimes (A^TP_1+P_1A)-2(L\otimes P_1BB^TP_1)]\xi\nonumber\\
&&-2 \sum_{i=1}^N[(\alpha_i(t)-\gamma-\|B^TP_1\| )\|r_i\| +\alpha_i(t)\beta]\nonumber\\
&&\bigg\|\sum_{j\in \mathcal{N}_i}[B^TP_1(\xi_i(t)-\xi_j(t))]\bigg\|\nonumber\\
&&
+2\sum_{i=1}^N \widetilde{\alpha}_{i}(t)\vartheta_i\bigg\|\sum_{j\in \mathcal{N}_i}[B^TP_1(\xi_i(t){-}\xi_j(t))]\bigg\|\nonumber\\
&=& \xi^T[L\otimes (A^TP_1+P_1A)-2(L\otimes P_1BB^TP_1)]\xi\nonumber\\
&&-2 \sum_{i=1}^N[(\alpha-\gamma-\|B^TP_1\| )\|r_i\| +\alpha\beta]\nonumber\\
&&\bigg\|\sum_{j\in \mathcal{N}_i}[B^TP_1(\xi_i(t)-\xi_j(t))]\bigg\|.
\end{eqnarray}
Adaptively updating $\alpha>\gamma+\|B^TP_1\|>0$, and choosing $\beta>0$, one has
\begin{eqnarray}\label{AdddLV1}
\dot{ {V}}_3
&\leq& -\xi^T(L\otimes Q_1)\xi\triangleq-U(t)\leq 0,
\end{eqnarray}
which implies that $V_3(t)$ is non-increasing. Then, according
to (\ref{AV3}), it follows that $\xi, \alpha_i(t)$ are bounded.
It is following from Assumption 4 that $r$ is bounded.
One has $\|F(r, t)\|=\|F(r, t)-F(0, t)\|\leq \gamma \|r\|$, which implies that $F(r, t)$ is bounded.
Therefore, $\dot{\xi}$ is bounded, which implies that
$\lim_{t\to\infty} V_3 (t)$ exists and is finite. Since (\ref{AdddLV1}), one has
one has
$\int_0^\infty U(t)dt$ exists and is finite. By noting that $\dot{U}(t)$ is also bounded. Therefore, ${U}(t)$ is uniform continuity. By utilizing Barbalat's
Lemma, it guarantees
$\lim_{t\to\infty}U(t) = 0$.  Thus, one has $\lim_{t\to\infty}\xi(t) = 0$. Noting that $\chi>0, \beta>0$, one has $\alpha_i(t)$
is monotonically increasing and
bounded. Thus, $\alpha_i(t)$ converges to some finite constants. Thus, it follows that
$
\lim_{t\rightarrow \infty}\xi_i(t)=\lim_{t\rightarrow \infty}\bigg(p_i(t)-\frac{1}{N}\sum_{k=1}^{N}p_k(t)\bigg)
=0.
$\\
Second, similar to the proof in Theorem 1, one has
\begin{eqnarray}\label{AEAL referencesignals}
\dot{\zeta}(t)
&=&(A+BK_1){\zeta}(t)+\varpi(t),
\end{eqnarray}
where $\varpi(t)=\sum_{i=1}^N\alpha_i(t)\vartheta_ih\bigg(\sum_{j\in \mathcal{N}_i} K_1(p_i(t)-p_j(t))\bigg)$.
Note that $\alpha_i(t)$ converges to some finite constants. By leveraging input-to-state stability to analyze the system (\ref{AEAL referencesignals}), one has $\lim_{t\to\infty}{\zeta}(t) = 0$. Then, one has
$\lim_{t\rightarrow \infty}\bigg(p_i(t)-\frac{1}{N}\sum_{k=1}^{N}r_k(t)\bigg)=0$.\\
Third, consider the following Lyapunov candidate
\begin{eqnarray}\label{V4}
V_4=\tilde{x}^T(I\otimes P_2)\tilde{x}+\sum_{i=1}^{N}\frac{\widetilde{\mu}_{i}(t)^2}
{\kappa_i},
\end{eqnarray}
where $\widetilde{\mu}_{i}(t)={\mu}_{i}(t)-{\mu}$.
As the proof given by Theorem 1, one has the derivation of (\ref{V4}) along (\ref{AMeagents_u}) and (\ref{Adaptive mu}),
\begin{eqnarray}\label{AddV2}
\dot{V}_4
&\leq&\tilde{x}^T(I\otimes (A^TP_2+P_2A-2P_2BB^TP_2))\tilde{x}\nonumber\\
&&-2\sum_{i=1}^N(({\mu}_{i}(t)-\gamma)\|x_i-r_i\|+{\mu}_{i}(t)\nu) \|B^TP_2\tilde{x}_i(t)\|\nonumber\\
&&+2\sum_{i=1}^{N} \widetilde{\mu}_{i}(t)\phi_i\|B^TP_2\widetilde{x}_i(t)\|\nonumber\\
&\leq&\tilde{x}^T(I\otimes (A^TP_2+P_2A-2P_2BB^TP_2))\tilde{x}\nonumber\\
&&-2\sum_{i=1}^N(({\mu}-\gamma)\|x_i-r_i\|+{\mu}\nu) \|B^TP_2\tilde{x}_i(t)\|
.
\end{eqnarray}
Adaptively updating  $\mu\geq \gamma$ and choosing $\nu>0$, one has
\begin{eqnarray}\label{AdddV2}
\dot{V}_4
&\leq&-\tilde{x}^T(I\otimes Q_2)\tilde{x}\triangleq -W(t)\leq 0,
\end{eqnarray}
which implies that $V_4(t)$ is non-increasing. Then, according
to (\ref{V4}), it follows that $\widetilde{x}, \mu_i(t)$ are bounded.
It is following from Assumption 4 and (\ref{A closedloop}) that $r$ and $p$ are bounded.
One has $\|F(x, t)-F(r, t)\|\leq \gamma\|x-r\|\leq \gamma (\|\widetilde{x}\|+\|p\|+\|r\|)$, which implies that $F(x, t)-F(r, t)$ is bounded.
Therefore, from (\ref{AMeagents_u}), one has
$\dot{\widetilde{x}}$ is bounded, which implies that
$\lim_{t\to\infty} V_4 (t)$ exists and is finite. Thus,
$\int_0^\infty W(t)dt$ exists and is finite. By noting that $\dot{W}(t)$ is also bounded. Therefore, ${W}(t)$ is uniform continuity. By utilizing Barbalat's
Lemma, it guarantees
$\lim_{t\to\infty}W(t) = 0$.  Thus, one has $\lim_{t\to\infty}\widetilde{x}(t) = 0$. Noting that $\kappa_i>0, \nu>0$, one has $\mu_i(t)$
is monotonically increasing and
bounded. Thus, $\mu_i(t)$ converges to some finite constants. It follows that
$
\lim_{t\rightarrow \infty}\widetilde{x}_i(t)=0$, which implies
$\lim_{t\rightarrow \infty}\bigg(x_i(t)-\frac{1}{N}\sum_{k=1}^{N}r_k(t)\bigg)
=0.
$
The proof is completed.

{\remark{Differing from the robust distributed average tracking algorithms (\ref{control input}) and (\ref{L distributed control algorithm}) in above subsection, the adaptive algorithms (\ref{Adaptive control input})-(\ref{Adaptive alpha}) are local fashion without knowing the global information $\gamma$.}}

\subsection{Continuous distributed average tracking algorithms design}
In the {above} subsections, the distributed average tracking algorithms are designed based on the discontinuous function $h(z)$, which may generate chattering phenomenon. In order to reduce the chattering in real applications
and make the controller easier to implement, based on the boundary layer
concept, we replace the discontinuous function $h(z)$ by a continuous approximation $h_\varepsilon(z)$, and propose a continuous distributed average tracking controller algorithm:
\begin{eqnarray}\label{Continuous Adaptive control input}
u_i(t) &=&K_1(p_i(t)-r_i(t))+K_2\tilde{x}_i(t)
+\mu\phi_ih_\varepsilon[K_2\tilde{x}_i(t)]\nonumber\\
&&+\alpha\vartheta_iBh_\varepsilon\bigg(\sum_{j\in \mathcal{N}_i} K_1(p_i(t)-p_j(t))\bigg),
\end{eqnarray}
and an continuous distributed average tracking filter algorithm
\begin{eqnarray}\label{Continuous Adaptive DAT algorithm}
{p}_i(t) &=& s_i(t)+r_i(t),\nonumber \\
\dot{s}_i(t) &=&As_i(t)+BK_1(p_i(t)-r_i(t))\nonumber\\
&&+\alpha \vartheta_iBh_\varepsilon\bigg(\sum_{j\in \mathcal{N}_i} K_1(p_i(t)-p_j(t))\bigg).
\end{eqnarray} 
Submitting (\ref{Continuous Adaptive control input}) into (\ref{agent}), one obtains the closed-loop systems in matrix form like:
\begin{eqnarray}\label{Continuous Adaptive control inputM}
\dot{\tilde{x}}(t)&=&(I\otimes(A+BK_2))\tilde{x}(t)
+(I\otimes B)(F(x,t)-F(r,t))\nonumber\\
&&+ (\mu \Phi\otimes B) H_\varepsilon[(I\otimes K_2)\tilde{x}(t)].
\end{eqnarray}
 It follows from (\ref{L referencesignals}) and (\ref{Continuous Adaptive DAT algorithm}) that
\begin{eqnarray}\label{continuous AM closedloop}
\dot{\xi}(t)&=&(I\otimes (A+BK_1))\xi(t)-(M\otimes BK_1)r(t)\nonumber\\
&+&(M\otimes B)F(r,t)+(\alpha M \Theta\otimes B) H_\varepsilon((L\otimes K_1)\xi(t)).
\end{eqnarray}

\begin{theorem}\label{Continuous A theorem1f}
Under Assumptions 1-4, the adaptive DAT algorithms (\ref{Continuous Adaptive control input}) and (\ref{Continuous Adaptive DAT algorithm}) solve the DAT problem of the multi-agent system (\ref{agent}) with the reference dynamical system (\ref{L referencesignals}) if the parameters are given by Algorithm 1.
\end{theorem}
\textbf{Proof}: First, consider the Lyapunov candidate (\ref{AV3}).
As proved in Theorem 1, the derivation of (\ref{AV3}) along (\ref{continuous AM closedloop}) is given by
\begin{eqnarray}\label{AddLV1}
\dot{ {V}}_1
&\leq&  \xi^T[L\otimes (A^TP_1+P_1A)-2(L\otimes P_1BB^TP_1)]\xi\nonumber\\
&&+2 \sum_{i=1}^N[(\gamma+\|B^TP_1\| )\|r_i\|]\bigg\|\sum_{j\in \mathcal{N}_i}[B^TP_1(\xi_i(t)-\xi_j(t))]\bigg\|\nonumber\\
&&-2\sum_{i=1}^N{\alpha}\vartheta_i\bigg(
\sum_{j\in \mathcal{N}_i}K_1(\xi_i(t){-}\xi_j(t))\bigg)\nonumber\\
&&h_\varepsilon\bigg(
\sum_{j\in \mathcal{N}_i}K_1(\xi_i(t){-}\xi_j(t))\bigg).
\end{eqnarray}
Since $\alpha>\gamma+\|B^TP_1\|$ and $\beta>0$, one has
\begin{eqnarray}\label{aaa}
\dot{V}_1(t)&\leq&  \xi^T[L\otimes (A^TP_1+P_1A)-2(L\otimes P_1BB^TP_1)]\xi\nonumber\\
&&+2 \sum_{i=1}^N\alpha\vartheta_i\bigg[\bigg\|\sum_{j\in \mathcal{N}_i}[B^TP_1(\xi_i(t)-\xi_j(t))]\bigg\|\nonumber\\
&&-\bigg(
\sum_{j\in \mathcal{N}_i}K_1(\xi_i(t){-}\xi_j(t))\bigg)\nonumber\\
&&h_\varepsilon\bigg(
\sum_{j\in \mathcal{N}_i}K_1(\xi_i(t){-}\xi_j(t))
\bigg)\bigg] \nonumber\\
&\leq&
-\eta V_1+2 \sum_{i=1}^N\alpha\vartheta_i\varepsilon e^{-ct}. \end{eqnarray}
In light of the well-known Comparison Lemma, one gets that
\begin{eqnarray}
{V}_1(t)
&\leq&
e^{-\eta(t)} V_1(0) +2 \sum_{i=1}^N\alpha\overline{\vartheta}_i
\int_0^t\varepsilon e^{-\eta(t-\tau)-c\tau}d\tau,
\end{eqnarray}
where $\overline{\vartheta}_i$ is the supper bound of ${\vartheta}_i$. According to
$\lim_{t\to \infty}\int_0^t\varepsilon
e^{-\eta(t-\tau)-c\tau}d\tau=0$, one has
$V_1(t)$ exponentially converges to the origin
as $t\to \infty$.
Therefore, $\lim_{t\to \infty}\|p_i-\sum_{k=1}^Np_k\|=0$.
Second, similar to Theorem 1, one has
\begin{eqnarray}\label{Continuous AEAL referencesignals}
\dot{\zeta}(t)
&=&(A+BK_1){\zeta}(t)+\varpi(t, \varepsilon),
\end{eqnarray}
where $\varpi(t, \varepsilon)=\sum_{i=1}^N\alpha\vartheta_ih_\varepsilon\bigg(\sum_{j\in \mathcal{N}_i} K_1(p_i(t)-p_j(t))\bigg)$. Since $\lim_{t\to \infty} \sum_{j\in \mathcal{N}_i} K_1(p_i(t)-p_j(t))=0$. One has $\lim_{t\to \infty} \varpi(t, \varepsilon)=0$. It follows that $\lim_{t\to \infty} {\zeta}(t)=0$. Thus, $\lim_{t\to \infty}\|p_i-\sum_{k=1}^Nr_k\|=0$.
Third, consider derivative of $V_2$ along (\ref{Continuous Adaptive control inputM}), one gets \begin{eqnarray}\label{Continuous ddV2}
\dot{V}_2
&\leq&\tilde{x}^T(I\otimes (A^TP_2+P_2A-2P_2BB^TP_2))\tilde{x}\nonumber\\
&&
+2\sum_{i=1}^N\|B^TP_2\tilde{x}_i(t)\|\gamma\|x_i-r_i\|\nonumber\\
&&-2\mu\sum_{i=1}^N\phi_i (B^TP_2\tilde{x}_i(t))^Th_\varepsilon(B^TP_2\tilde{x}_i)\nonumber\\
&\leq&-\eta_2V_2+2\sum_{i=1}^N\mu\phi_i \varepsilon e^{-ct}.
\end{eqnarray}
Thus, $\lim_{t\to \infty}V_2(t)=0$, which implies  $\lim_{t\to \infty} \|x_i(t)-p_i(t)\|=0$. Thus, $\lim_{t\to \infty}\|x_i-\sum_{k=1}^Nr_k\|=0$.
This completes the proof.
\section{Conclusions}
In this paper, we have studied the distributed average tracking problem of multiple time-varying signals generated by nonlinear dynamical systems. In {the} distributed fashion, a pair of discontinuous algorithms with static and adaptive coupling strengths have been developed. Then, in light of the boundary layer concept, a continuous algorithm is designed. Besides, sufficient conditions for the existence of distributed algorithms are given. The future topic will be focused on the distributed average tracking problem for the case with only the relative output information of neighboring agents.

\bibliographystyle{plain}        


\end{document}